**Emergence of Specialization from Global Optimizing Evolution in a Multi-Agent System**

Zengru Di, Jiawei Chen, Yougui Wang, and Zhangang Han

*Department of Systems Science, Beijing Normal University, Beijing, 100875, China*

**ABSTRACT**

The evolution of specialization in a multi-agent system is studied both by computer simulation and Markov process model. Many individual agents search for and exploit resources to get global optimization in an environment without complete information. With the selection acting on agent specialization at the level of system and under the condition of increasing returns, the division of labor emerges as the results of long-term optimizing evolution. Mathematical analysis gives the optimum division of agents and a Markov chain model is proposed to describe the evolutionary dynamics. The results are in good agreement with that of simulation model.

## Ⅰ. INTRODUCTION

The emergence of collective behavior in multi-agent systems has been an interesting area of complexity research. Multi-agent systems are characterized by vast number of agents trying to gain access to limited resources in an unpredictable environment. This description applies to a wide range of systems, ranging from ecology, economy to computer networks [1-6]. In a distributed multi-agent system, be it natural or social, agents usually do not have complete information about the system in which they are embedded, and at the same time they must successfully interact with each other in order to get global optimization. Although the interactions among agents are simple and local, it can lead to complex dynamics at the global scale. Studies of computational ecology have shown that when the agents make choices in the presence of delayed and imperfect knowledge about the state of the system, their dynamics could give rise to nonlinear oscillations, clustered volatilities and chaos that drive the system far from optimality [7, 8]. In natural and social systems for example in the ecologically important social insects, the colony is self-organized as an integrated unit. The agents in the system have a hierarchical organization that determines the partitioning of reproduction, resources, and tasks. So the colony as a whole is able to manage a complex and changing foraging area [9, 10]. Holland has argued that these complexities arise from the self-adaptive properties of the individual agent [11]. The approach of complex adaptive systems has been applied to a wide range of systems, including biological ecosystems and economic systems [12-14].

Specialization or division of labor observed in many complex systems is one of the most striking examples of collective behavior. Roughly speaking, an economic organizational pattern is said to involve division of labor if it allocates labor of different individuals to different activities. Hence the specialization of individuals and the number of professional activities are the two sides of division of labor [15]. Division of labor is a fundamental way to improve efficiency and utilization so as to get global optimization for the system. In social insects colony, the most obvious sign of the division of labor is the existence of castes. The individuals belonging to different castes are usually specialized for the performance of a series of precise tasks. A lot of works have been done to study the formation of division of labor and the mechanism for tasks allocation [10, 16-19]. In order to understand the mechanisms behind the formation of specialization, evolutionary processes and principles are helpful. From the viewpoint of long-term evolution, selection acting on agent specialization must take place at the level of the colony. Some colonies survive and reproduce more than others because they have a division of labor that is better adapted for a particular environment[20, 21]. Actually, the evolutionary processes and principles play a fundamental role for the development of both natural and man-made systems. This includes molecular, genetic or cellular level as well as ecological, economic, social and technological problems. Despite the diversity of time and space scales involved all these processes are governed by the same principles of competition, mutation and selection. In this paper, a simple model for studying the evolutionary process of specialization is developed. The model describes a system of many individual agents that search for and exploit resources to get global optimization in an environment without complete information. There are two kinds of tasks for every agent: searching in the area to find unknown resources or exploiting the resource that is known as the best one to the agent. The behavior character of every agent - that is the probability for the agent to search for or to exploit resources - is described by a real number and it can be changed in the process of evolution. In each period we reward agents according to their actual performance to the global optimization of the system. An algorithm is given to describe the genetic variation and natural selection. As well shall see, with long-term evolution the system usually forms a certain macroscopic structure. Under the condition of increasing returns, specialization in deterministic exploiting and stochastic searching behavior is always the results from the global optimizing evolution from certain given initial conditions, such as the agents are all homogenous, random or uniform distributed. The results reveal that the random mutation in evolutionary process is necessary to form macroscopic structure and to reach global optimum. Meanwhile the stochastic behavior of some agents is meaningful for the system to deal with uncertain environments. A mathematical analysis based on Markov chain process gives similar results. Our work provides several insights that are useful knowledge for us to understanding the evolutionary dynamics in biology and the formation of organization.

The presentation is organized into two major parts. In Section II the model for computer simulation is presented. Resources distribution and related growth dynamics are qualified. Agent's behaviors, total returns of all agents as fitness function of the system, and algorithm for evolutionary process are also given. Then the computer simulation results are reported. In Section III, capturing the basic features of the above simulation model, we present a simpler situation for mathematical analysis. We show that when there is increasing returns in agents' behavior, the division of agents in searching and exploiting is a necessary condition for global optimization. A Markov chain process model to describe the evolutionary dynamics of the system is introduced. The corresponding results of mathematical and simulation model are consistent well. In Section IV, we provide a summary of our results and a brief discussion of some unresolved issues that remain.

## II. SPATIAL EVOLUTIONARY MODEL AND SIMULATED RESULTS

Consider a system with $M$ individual agents. Every agent is a highly autonomous entity who can search for and exploit resources in a given environment. There is complete information transfer among agents. Each agent knows the information about the resource distribution foraged by every other agent. The space and resources distributions, agent properties and rules for evolutionary dynamics are defined as follows.

### A. Spatial evolutionary model

1. Environment and resource distribution

The space for the colony to live is a discrete $L \times L$ lattice of patches. The resources are random distributed in the lattice as a several of isolated regions. Its value at point $i$ is denoted as $S(i)$. The time is also discrete changed and the growth of resource is determined by Logistic equation with a proportional exploiting if there is an agent

$$S_t(i) = aS_{t-1}(i)(1 - \frac{S_{t-1}(i)}{M(i)}) - bS_{t-1}(i). \tag{1}$$

For the simplicity we have used the fixed boundary condition in the simulation.

2. Agents properties

$M$ agents are randomly scattered over the lattice in the beginning of simulation. Every agent has only local knowledge about the resource distribution. The agent only knows the size of resources of its nearest 8 neighbors. But every agent knows the resources size of any other agent knows. Let's denotes these 8×*M* (or less) lattice points as set *Φ*. The agent can move to, search for and exploit resources in any point that is not occupied by other agent over the lattice. In any time period if an agent is in the lattice point *i* with resource size *S*(*i*), its product is *F*=*bS*(*i*). When the agent moves, the cost *C* is determined by a coefficient *β* and the distance between starting point *i* and end *j*:

$$C = \beta d(i, j) = \beta \sqrt{(x(i) - x(j))^2 + (y(i) - y(j))^2} \ . \tag{2}$$

In the beginning of each time period, every agent would evaluate the information of resources distribution to find the best point that he can get maximum return. Suppose an agent *k* is at lattice point *i*, his best point is determined by:

$$\underset{j \in \Phi}{Max}[b(S(j) - S(i)) - \beta d_{ij}] . \tag{3}$$

A real number $q_k$ defined in [0, 1] is used to describe the behavior character of agent *k*. It stands for the probability of random move of the agent. On the basis of above evaluation, the agent *k* has a probability $q_k$ to random move over the lattice and a probability 1- $q_k$ to go to the point with maximum return. In each time period *t*, the total return *R* of this multi-agent system is the sum of net income of every agent:

$$R_t = \sum_{k=1}^{M} r_t^k = \sum_{k=1}^{M} (F_t^k - C_t^k) = \sum_{k=1}^{M} (bS_t^k - \beta d_t^k) . \tag{4}$$

In our computer simulation, for simplicity and without loosing any generality, the above decision making procedure does not proceed in a parallel way. Every agent makes his decision in proper order.

Parameter *q* has determined the behavioral character of agent. An agent with *q* near 0 tends to get maximum return from the known resources and on the contrary an agent with *q* near 1 tends to move randomly over the lattice that is helpful to discover the unknown resources. From the effect of "learning by doing" or increasing returns discussed frequently in economics [22, 23], we assume that the characteristic parameter $q_k$ is not only determining the movement of agent *k*, but also related to the efficiency of resources exploiting and searching. That is:

$$\begin{aligned} b_k &= b_0(1 - q_k) \\ \beta_k &= \beta_0(1 - q_k) \end{aligned}, \tag{5}$$

The goal of the system is to maximize the global return for a generation with *N* periods, that is $W = Max \sum_{t=1}^{N} R_t = NM\bar{r}$, where $\bar{r}$ is the average returns of each agent in one time period of the generation. From the assumption mentioned above, especially the assumption of global optimization and perfect information among agents, we know that the multi-agent system has already formed an aggregate unit. Then the major problem here is that what distribution of the agents (the distribution of $q_k$) could let the colony get global optimization and how can the system reach its optimum distribution through evolutionary process.

3. Mutation, replication and natural selection

Based on the processes believed to operate in biological evolution, the evolutionary search of the system is defined by the following steps.

As an initial condition, generate a system with $M$ agents and any given initial distribution $q_k$.

(1) Run the model for $N$ periods as a generation. Calculate $\bar{r}$ -the average returns of each agent in one period of this generation. Calculate the average products $\bar{F}$ and the average moving cost $\bar{C}$. Using $\bar{F}$ and $\bar{C}$, we divide all agents into three subgroups, that is:

$$\Psi_0 = \{\text{Agent } k \mid F_k \geq \bar{F} \text{ and } C_k \leq \bar{C}\},$$
$$\Psi_2 = \{\text{Agent } k \mid F_k < \bar{F} \text{ and } C_k > \bar{C}\}, \tag{6}$$

and $\Psi_1$ includes all the other agents.

(2) Mutation. Create a new generation by varying the characteristic parameter $q_k$ of every agent with probability $p_1$. If the character of agent $k$ varies, it has the same probability ($p=0.5$) for changing to $q_k+\Delta q$ or $q_k-\Delta q$. If $q_k+\Delta q$ is bigger than 1, then let $q_k$ equal 1. If $q_k-\Delta q$ is smaller than 0, then $q_k$ equals 0. Run the model as step 1. Get the average returns of each agent in one time period of new generation $\bar{r}_{new}$

(3) Replication and natural selection.

If $\bar{r}_{new}$ is bigger than $\bar{r}$ of last generation, that means the mutation is helpful to the global optimization of the system, then replicate and strengthen the variant. Replication is to create another agent with the same characteristic parameter as the variant and replace an unchanged agent in subgroup $\Psi_2$ randomly to keep the size of the system unchanged. Strengthen is adding another $\Delta q$ to the characteristic parameter of the variant in the same direction of variation. That is if the parameter of variant is $q_k+\Delta q$, then in the new generation its parameter is $q_k+2\Delta q$, and if the parameter of variant is $q_k-\Delta q$, then in the new generation its parameter is $q_k-2\Delta q$.

If $\bar{r}_{new}$ is smaller than $\bar{r}$ of last generation, that means the mutation is not helpful to the global optimization of the system, then replacing the variant with probability $p_2$ by the agent in subgroup $\Psi_0$ randomly.

For the new generation formed by replication and natural selection, run model as step (1).

(4) Go to step 2.

In the above genetic variation and natural selection, the mutation only changes the character of agent gradually. This is what happened in real biological system. In fact, any obvious diversion and bifurcation happened in biological system is the accumulation of effects of gradually variation in the long-term evolutionary process. So our mutation mechanism is rationale. In section Ⅳ other mechanisms for mutation and selection are discussed. Another point should be mentioned is that because the evaluation for the variant is based on its effect on total returns of the system, the selection here has the same effect as it takes place at the level of colony.

**B. Simulation results**

The simulations are typically done in 100×100 2-D lattice with 8 isolated resource regions. 30 agents form a colony and try to gain access to distributed resources. The other parameters are: $a$=0.05, $b_0$=0.5, $\beta_0$=0.6, $p_1$=0.1, $p_2$=0.5, and character changing when mutation happens $\Delta q$=0.05. The character space [0,1] is divided into 20 smaller intervals by $\Delta q$. The number of agents at each interval (denoted by $N(q)$) gives the distribution of agents. The typical behavior of the evolution is the increasing of global returns (described by $\bar{r}$) with the process of specialization. Figure 1 are the results from homogeneous initial conditions, that is every agent has the same character parameter: $q_k$=0.52 (for $k$=1 to $M$). Figure 1(a) gives the epochal evolution of the distribution of number of agents $N(q)$ represented by character $q$. The division of agents in different tasks occurs as the result of long- term evolution. Fig. 1(b) shows the changing of $\bar{r}$ with the evolution.

(FIGURE 1)

In all of our simulations, $\bar{r}$ rises with the evolution of colony. For some initial conditions, $\bar{r}$ goes through stepwise changes. That is typically observed in some optimizing epochal evolutionary search [24]. Later in the evolution, $\bar{r}$ reaches its optimal "steady state" with higher fluctuations. The average distribution of number of agents $N(q)$ in the generations of optimal stasis is shown in Figure 1(c). In the macroscopic level, division of labor emerges with the global optimizing evolutionary process.

(FIGURE 2)

Figure 2 shows the evolution behavior with two different initial distributions. Figure 2(a) is the behavior with random initial distribution and Figure 2(b) is the situation that every agent has the same character parameter $q$=0.82 in the initial. It could be seen that the finial results (the average distribution of number of agents $N(q)$ and the average optimal returns) are almost not related to the initial distributions.

We have also studied the evolutionary behavior with other parameter settings. Figure 3 are the results of 4 isolated resource regions randomly distributed in 100×100 2-D lattice. All the other parameters are the same as in Figure 1. Comparing these two simulation results, we could find that the less obtainable resources in a given space leads to the more agents specialized in stochastic searching behavior. This is rational and is consistent with the results of the mathematical analysis in the next section.

(FIGURE 3)

It should be mentioned that our evolutionary model has a problem in its dynamical simulations. That is some times the colony could not maintain the agents that specialized in random search. The reason may be in the mechanism for genetic variation and natural selection. When the random searching agent ($q$=1) happens to vary in the direction of decreasing $q$, it could lead to the decreasing of global returns. And then this agent would be replaced by the agent in subgroup $\Psi_0$, which is normally composed by the deterministic exploiting agent with $q$ around 0. The related problem is that when all the agents are distributed near $q$=0, it is very difficult to reach the optimal distribution with specialization. The other reasons for this problem would be related to the parameters of our computer simulation. We guess that if the simulation proceeds with larger space, more agents, and less character changing for mutation ($\Delta q$), the situation would be better. In the next section, we haven't found the above problem in the corresponding mathematical model.

## III. MATHEMATICAL ANALYSIS FOR THE GLOBAL OPTIMIZING EVOLUTION

Without losing any generality, we introduce a simple, non-spatial model of evolutionary dynamics, so that some mathematical analytical results can be achieved. And the main results observed in simulations have been reproduced.

### A. Optimal behavior for solitary agent

At first, let's discuss the problem for a solitary agent to deal with an uncertain environment.

The living space for the agent is composed by the samples of resources valued $F_0$ and 0. In the beginning of every period, the agent can choose to search for new resources (with probability $q$) or to take the situation of last period (with probability $1-q$). When the agent determines to search new resource, it will get the resource $F_0$ with probability $P$. The corresponding product is $(1-q)F_0$. The agent has searched for nothing (valued 0) with probability $1-P$. The cost for the search is $(1-q)C$. The factor $(1-q)$ describes the effect of learning by doing. When the agent determine to take the last situation, if the last value of the resource is $F_{t-1}$, the value of this period is

$$F_t = \frac{1}{a} F_{t-1}, \tag{7}$$

where $a>1$ is a parameter related to the diminishing of resource under exploiting. The goal of the agent is to maximize total net returns for $N$ periods. That is determined by parameter $q$ for a given distribution of resources (described by probability $P$).

In the beginning of searching and exploiting process noted as period 0, the agent proceeds with a random search. For this period, the expected return is $E_0 = (1-q)(PF_0 - C)$. For period 1, the agent can search for new resources with probability $q$. The corresponding expected benefit is $E_0$. Another choice is stay in the situation of last period (with probability $1-q$). The corresponding expected return is $PF_0/a$. So the expected return of period 1 is

$$E_1 = (1-q)(qE_0 + (1-q)\frac{1}{a}PF_0). \tag{8}$$

From the similar analysis, we could get the expected return for the period 2:

$$E_2 = (1-q)(qE_0 + (1-q)q\frac{1}{a}PF_0 + (1-q)(1-q)\frac{1}{a^2}PF_0). \tag{9}$$

Let $x=(1-q)/a$, proceed the same analysis, the expected benefit of period $n$ can be written as

$$\begin{aligned} E_n &= (1-q)(qE_0 + xqPF_0 + x^2 qPF_0 + x^3 qPF_0 + \ldots + x^n PF_0) \\ &= (1-q)(qE_0 + PF_0(q(x+x^2+x^3+\ldots x^{n-1}) + x^n)) \\ &= (1-q)(qE_0 + PF_0(x^n + q\frac{x(1-x^{n-1})}{1-x})) \end{aligned} \tag{10}$$

Hence the total expected returns from period 0 to period $N$ is

$$\begin{aligned} W &= \sum_{n=0}^{N} E_n = (1+N)(1-q)qE_0 + PF_0(1-q)\sum_{n=0}^{N}(x^n + q\frac{x(1-x^{n-1})}{1-x}) \\ &= (1+N)(1-q)qE_0 + PF_0(1-q)[(1+N)\frac{qx}{1-x} + (1-\frac{q}{1-x})\frac{1+x^N}{1-x}] \end{aligned} \tag{11}$$

From the above result, we can determine the optimal point $q_M$ and corresponding maximum total expected returns $W$. We can also find the relationship between $q_M$ and other parameters, such as searching cost $C$, probability $P$, number of period $N$, and parameter $a$. The results are given in Figure 4. Larger $P$, $N$, $a$ and smaller $C$ will lead to bigger searching probability $q$.

(FIGURE 4)

## B. Optimal division of agents for a colony

Let's now turn to the situation of a colony consist of $M$ agents. Each agent is characterized by a parameter $q_i$, which determines the searching probability of the agent. As discussed in the above solitary agent situation, we assume that every searching agent could find resources $F_0$ with probability $P$. The same as the assumption in the previous simulation models, there is also complete information among all agents. So if any agent has found a resource $F_0$, all the others will go and exploit it. And then the product of the system is $\sum_{i=1}^{M}(1-q_i)F_0$. The searching cost for each agent is $(1-q_i)C$.

Let's compare the results on the total returns of distributed $q_i$ with complete specialization. For a multi-agent system with distributed $q_i$, at every period and on the average there will be $\sum_{i=1}^{M} q_i = m$ agents searching new resources. So the gross probability of finding at least one new resource $F_0$ is the same as when there is $m$ agents specialized in searching. We denote this gross probability as $P_r$. For a colony with distributed $q_i$, the expected product of the first period is $R_D = \sum_{i=1}^{M}(1-q_i)F_0 P_r$
$= MF_0 P_r - F_0 P_r \sum_{i=1}^{M} q_i = F_0 P_r (M-m)$. The cost for searching is $C_D = \sum_{i=1}^{M} q_i(1-q_i)C$
$= mC - C\sum_{i=1}^{M} q_i^2$. With $q_i \leq 1$, $\sum_{i=1}^{M} q^2$ is usually less than $m$. So we usually have $C_D>0$. But for the colony with specialized agents, although the expected product of the first period is the same: $R_S = \sum_{i=1}^{M-m} F_0 P_r = F_0 P(M-m) = R_D$, and so that the products of following generations are also the same, the cost for search is 0. So the net return of specialized system is bigger than that of the distributed one.

Assuming that there is $m$ agents specialized in searching in the colony, all the others specialized in exploiting the resource. In each period, If any searching agent has the probability $P$ to find the resource $F_0$, then the probability for $m$ agents at least find one resource $F_0$ is $P_r = 1-(1-P)^m$. So the expected returns of the colony for every period are:

$$E_0 = P_r(M-m)F_0,$$

$$E_1 = P_r(M-m)F_0 + P_r(1-P_r)\frac{1}{A}(M-m)F_0,$$

$$E_2 = P_r(M-m)F_0 + P_r(1-P_r)\frac{1}{A}(M-m)F_0 + P_r(1-P_r)^2\frac{1}{A^2}(M-m)F_0.$$

……

Let $x=(1-P_r)/A$, where parameter $A>1$ is also related to the diminishing of resource under $M$ agents' exploiting. The return of $n$th period can be written as

$$E_n = E_0\left(\frac{1-x^n}{1-x}\right). \tag{12}$$

So the total returns of $N$ periods is

$$W = \sum_{n=0}^{N-1} E_0\left(\frac{1-x^n}{1-x}\right) = \frac{E_0}{1-x}\left(N - x\left(\frac{1-x^N}{1-x}\right)\right)$$
$$= \frac{P_r(M-m)F_0}{1-x}\left(N - x\left(\frac{1-x^N}{1-x}\right)\right), \tag{13}$$

By equation (13) we can get the optimal point $m_0$ and its relationship with other parameters. It has been found that $m_0$ is sensitive to the probability $P$ (as shown in Figure 5(a)). Figure 5(b) is the simulation results of the previous simulation model with complete specialized agents. Theoretical and simulation results are in good agreement.

(FIGURE 5)

Compare the simulation results in Section 3 with the results of complete specialized system mentioned above, we could find that our evolution process have almost got the best returns from a given environment. For the space with 8 resources, the average return for each period of complete specialized colony is 37.01. The corresponding evolution result is 35.39(average of the later stasis periods). For the environment with 4 resources, the average return of complete specialized system is 12.01. The corresponding evolution result is 10.48(average of the later stasis periods).

**C. Markov chain model for the evolutionary dynamics**

Let's assume the character space of the agent has $k+1$ states corresponding to the searching probabilities described by parameter $q_j$, $j=0, 1, \cdots\cdots, k$, with $q_0=0$ and $q_k=1$. The distribution of agents in every state describes the situation of the colony in macroscopic level. Let's denote this distribution as $\{N_j, j=0, 1, \cdots\cdots, k\}$, and we have $\sum_{j=0}^{k} N_j = M$. Here $N_j$ is a positive real number instead of a positive integer. Then from the results in the above discussion, the total product $W$ and total cost $C_D$ of the colony in one generation with $N$ periods is

$$W = \frac{P_r(M - \sum_{j=0}^{k} N_j q_j)F_0}{1-x}(N - x(\frac{1-x^N}{1-x})), \qquad (14)$$

$$C_D = N \sum_{j=1}^{k} N_j q_j (1-q_j) C, \qquad (15)$$

where $x=(1-P_r)/a$, $P_r = 1-(1-P)^m$, and $m = \sum_{j=0}^{k} N_j q_j$. The total return of the system in one generation is $R(\{N_j\}) = W(\{N_j\}) - C_D(\{N_j\})$.

(FIGURE 6)

With the evolution process between two generations, the state of every agent would transit among all the $k+1$ states. From our computer simulation model, only the transition between the nearest neighbors can happen. A Markov chain process as shown in Figure 6 could describe the genetic variation and natural selection in the evolution. The dynamical behavior of this Markov chain is determined by following equations:

$$\begin{aligned}N_j(t+1) = N_j(t) + P(j+1 \to j)N_{j+1}(t) + P(j-1 \to j)N_{j-1}(t) \\ - P(j \to j-1)N_j(t) - P(j \to j+1)N_j(t)\end{aligned} \qquad (16)$$

For $j=0$ and $j=k$ we have:

$$N_0(t+1) = N_0(t) + P(1 \to 0)N_1(t) - P(0 \to 1)N_0(t), \qquad (17)$$

$$N_k(t+1) = N_k(t) + P(k-1 \to k)N_{k-1}(t) - P(k \to k-1)N_k(t). \qquad (18)$$

$P(i \to j)$ is the transition probability for an agent from state $i$ to state $j$. It is determined by the global optimization. Corresponding to the natural selection process described in the simulation model, that is the replication or replacement of a varying agent according to its result on total returns, the transition probability could be written as

$$P(i \to j) = \frac{1}{2}\mu[5 + 3\operatorname{sgn}(R(N_i - 1, N_j + 1) - R(N_i, N_j))], \tag{19}$$

where $\mu$ is a parameter related to the probability of mutation. From the above equations, we can get the results of optimal evolution of the system from any given initial conditions (as shown in Figure 7).

(FIGURE 7)

The Markov chain model is proceed under the following parameters. $M=30$ agents form a colony and want to get global optimization in an uncertainty environment. The probability $P$ for every searching agent to find resource $F_0=10$ is 0.7 and the searching cost is $C=8$. The other parameters are $\Delta q = q_{i+1} - q_i = 0.05$, $A=4$, $N=5$, and $\mu=0.02$. Given any initial condition, we could get the evolution of the agent distribution. The results are almost the same as the computer simulations in Section Ⅱ (The average returns are all scaled in order to have the similar quantities as that of simulation results).

In the end of Section Ⅱ, we have mentioned that our computer simulation model some times could not maintain the agents that specialized in random search. This problem does not happen in the mathematical model. The final optimal distribution in the Markov chain process is stable and it could be achieved from any given initial conditions include that all the agents are initially homogeneous with $q_i=0.9$, $i=1$ to 30 (See Figure 8(a)). We have compared the final stable distribution of the Markov chain process with the average distribution in the generations of optimal stasis in Section Ⅱ. As shown in Figure 9, it is notable that the mathematical results are consistent well with the results of simulations.

(FIGURE 8)

(FIGURE 9)

### Ⅳ. CONCLUDING REMARKS

In this paper we have studied the formation of specialization by means of a simple multi-agent model with optimizing evolution and a few mathematical arguments. The main problem under consideration was the origins of specialization and its evolutionary dynamics. In summary our findings are the following.

1. Specialization is a kind of functional structure in macroscopic level of multi-agent systems.

The model demonstrated how a global structure could be generated in simulation. The division of agents in different tasks could emerge from the long-term optimizing evolution under the mechanism of increasing returns. The methods presented here also provide a way to study the emergence of other global structure of complex system through higher dimensional state space.

2. An evolutionary process is presented in this paper to describe the mechanism of mutation and natural selection. It gives some results on the evolutionary dynamics in multi-agent systems, including the emergence of optimal distribution. This optimizing evolutionary search method is also useful in the domains of optimization, adaptation and learning.

3. The results reveal that the stochastic properties in evolutionary process are necessary to generate macroscopic structure and to reach global optimization. Meanwhile the stochastic behaviors of some agents are meaningful for the colony to deal with uncertain environments.

This work suggests a number of future directions for the study of multi-agent systems. As was mentioned in section Ⅱ, the model assumes that there is perfect information among agents and the goal of the system is global optimization. These assumptions in fact have suggested that the agents have already formed a colony. It would be extremely interesting to study the mechanism of aggregation of individual agents. That is to study the relationship between individual behavior and global optimization, to see how organization emerges from individual optimum, and to understand how could a multi-agent system to form an aggregate units. Another problem is related to the mechanism of mutation and natural selection. We have also tried other mechanisms for evolution, such as when the mutation happens, the parameter $q$ could randomly take the value over [0, 1]; the colony is not divided into subgroups, whole invariant could replicate the worse variant and so on. But the mechanism we have presented here seems be the better one. Anyway, other mechanisms should be studied with the formation of aggregate unit in the evolution of multi-agent systems.

Several other interesting issues remains to be explored. Specialization and organization are typical characters in economic systems. Our method here gives an approach to understand the mechanism behind these innovation phenomena. It could give a dynamical perspective on the formation of economic structure. Because the division in high dimensional state space gives a nice description on diversity, our model also provides a valuable way to understand the diversities in biological and ecological systems. The Markov chain model in Section Ⅲ is a kind of coupled map lattice in discrete dynamical systems. It gives us some insights on the formation of macroscopic structure. It is also useful to study certain temporal-spatial properties of this dynamical system. By using this mathematical model and some parameters to describe the effect of increasing returns, we can study the emergence of structures in detail. Some preliminary results reveal that the functional structure emerges from a series of bifurcations.

## ACKNOWLEDGMENTS

The authors thank Dr. Qiang Yuan and Dr. Yiming Ding for helpful discussions. This work was supported by the State Natural Science foundation of China under grant No.79990580 and No.60003018.

**Figure Captions**

FIG.1. Typical evolutionary behavior of the system. Parameters are given in the beginning of section Ⅱ.B. (a) The vertical axes shows the distribution of number of agents $N(q)$, the evolution is measured in generations. (b) The average return $\bar{r}$ in the agents as a function of the number of generations. (c) The average distribution after the system reaching the optimal stasis.

FIG.2. Evolutionary behavior with different initial distributions. The plots show the distribution evolution and average returns as a function of generations. (a) Random initial distribution. (b) Homogeneous initial distribution with $q_k$=0.82 for $k$=1 to 30. There is stepwise increasing in average returns.

FIG.3. Simulation results with 4 isolated resource regions. The three plots show the distribution evolution (a), the evolution of average returns (b), and average distribution of optimal stasis (c).

FIG. 4 Total expected returns $W$ as a function of searching probability $q$ (horizontal axis) for solitary agent. The resource value $F_0$=20, the probability $P$=0.2 are the same for all curves. Other parameters are labeled near the corresponding curve.

FIG. 5 Returns (normalized) as a function of number of searching agent $m$. (a) Theoretical results. Except parameter $P$ (labeled above the corresponding curve), all the other parameters for each curve are the same. (b) Simulation results of a complete specialized colony. Less number of resources is corresponding to the smaller probability $P$ of finding new resources for searching agent.

FIG. 6. Markov chain process

FIG. 7 Evolution of the distribution of agents and average returns. Compare with the simulation results in Fig. 1 and Fig. 2, the results are almost the same. But in our Markov chain model, there is no longer stepwise increasing in average returns (Compare (b)2 here with Fig. 2 (b)2).

FIG. 8 Evolution from different initial conditions. (a) Homogeneous initial distribution with $q_k$=0.1 for every agent. (b) Random initial distribution.

FIG.9 Comparison of theoretical predictions (solid lines) and simulation results (column bars) of the distributions of agents' number. In the graph, $N(0)$ and $N(1)$ of the theoretical results have been corresponding to the number of agents at the interval [0, 0.05] and [0.95, 1] in simulation. (a) $P$=0.7 in mathematical model with 8 resources in simulations. (b) $P$=0.4 in mathematical model with 4 resources in simulations.

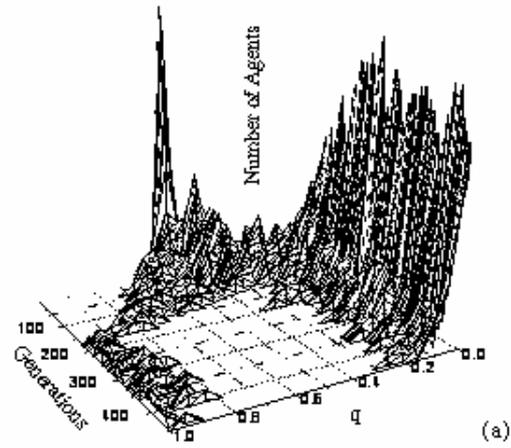

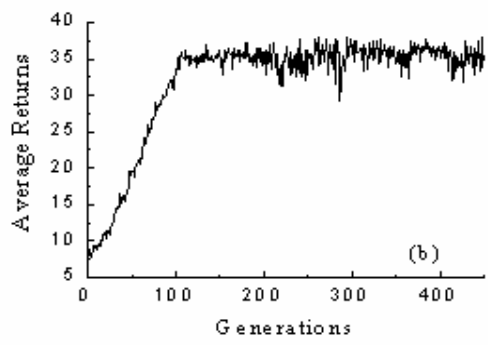

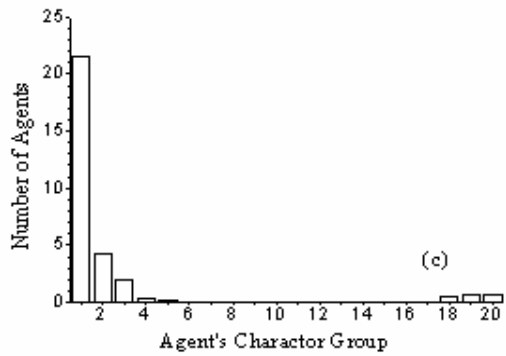

Figure 1—Di: Specialization

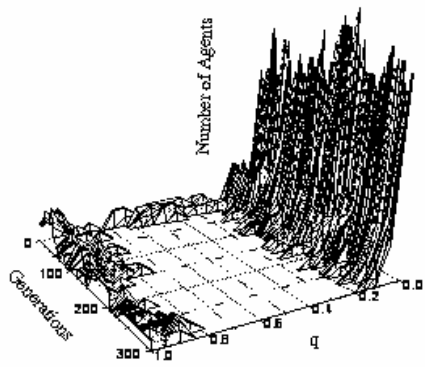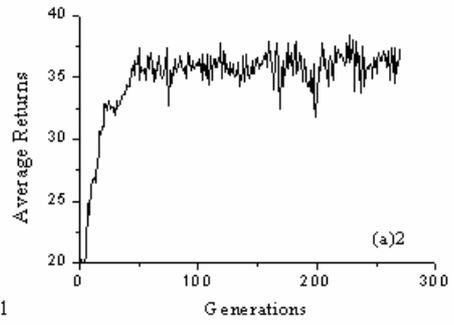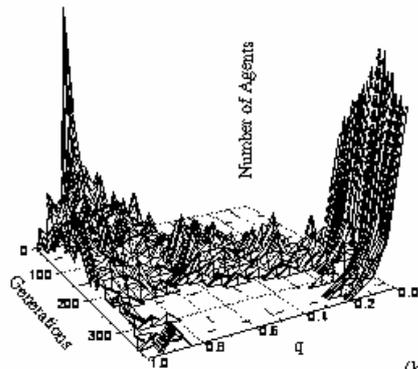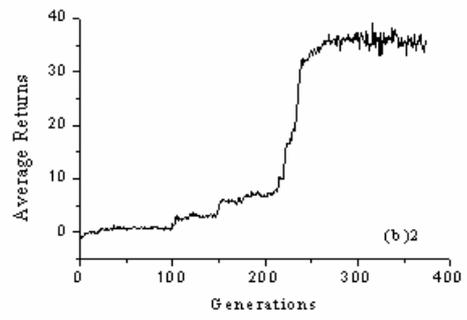

Figure 2—Di: Specialization

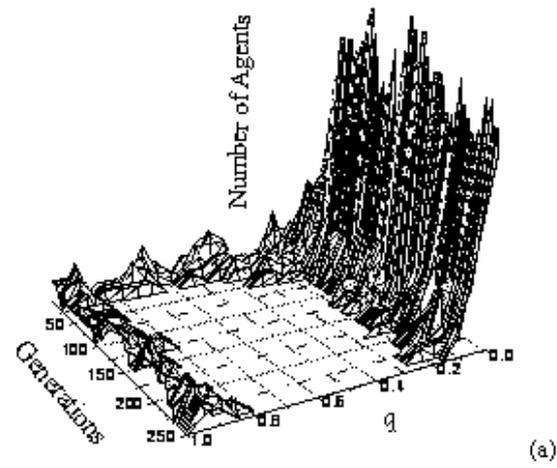
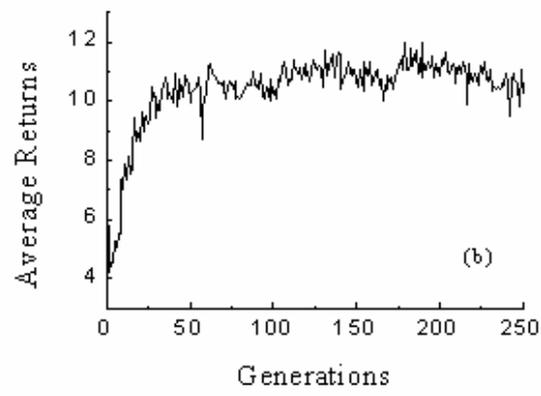
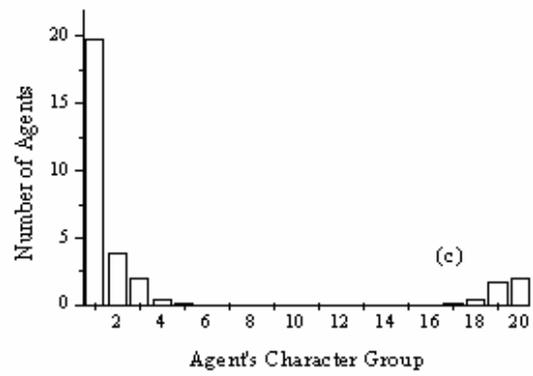

Figure 3—Di: Specialization

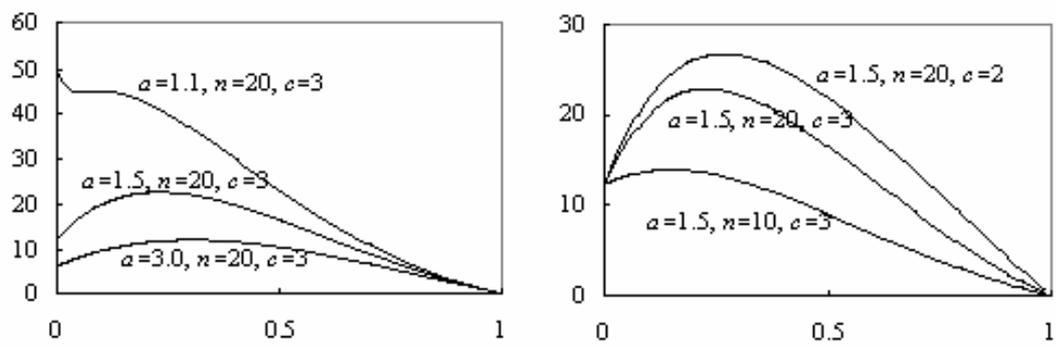

Figure 4—Di: Specialization

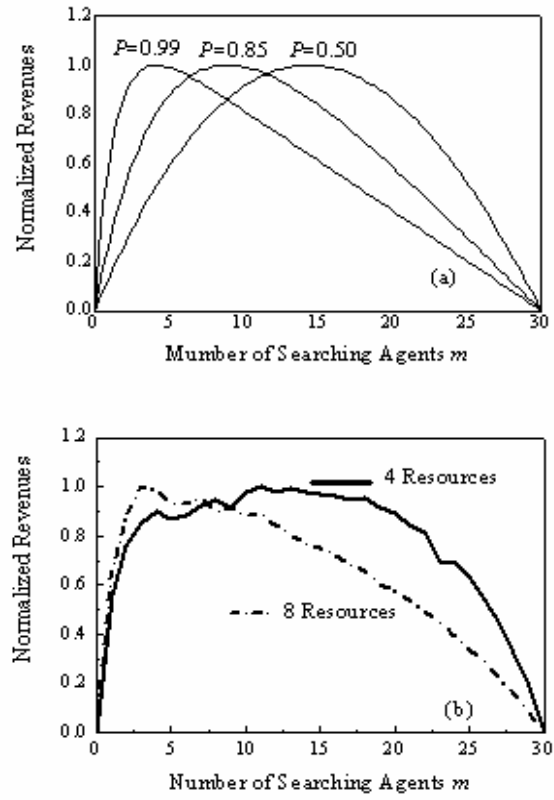

Figure 5—Di: Specialization

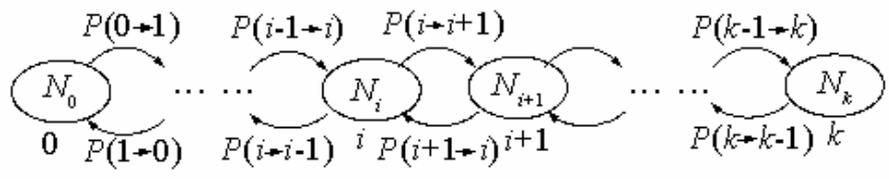

Figure 6—Di: Specialization

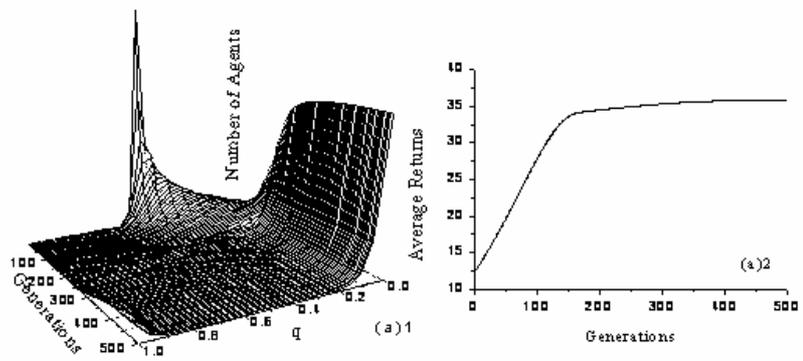

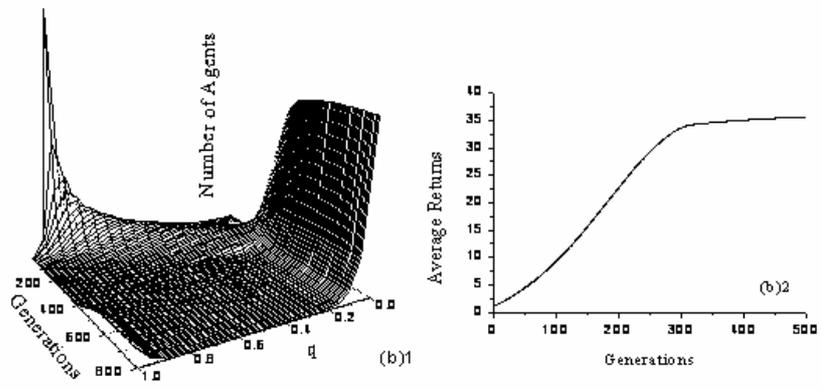

Figure 7—Di: Specialization

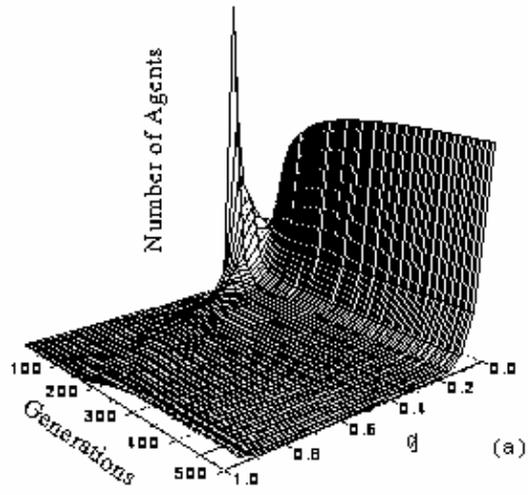

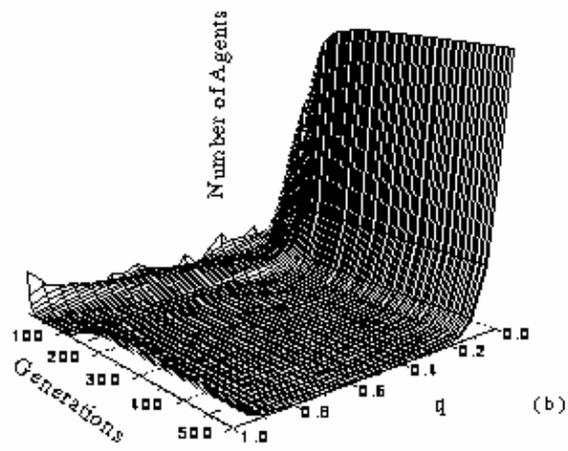

Figure 8—Di: Specialization

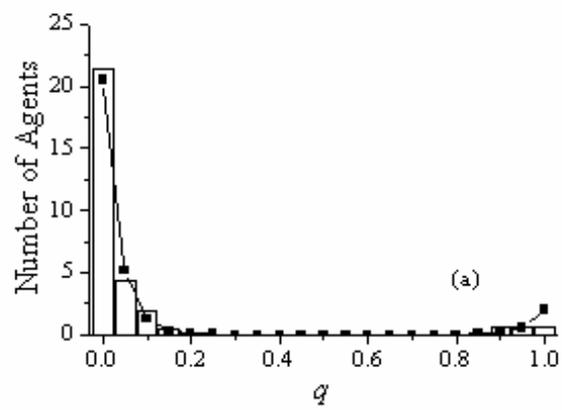

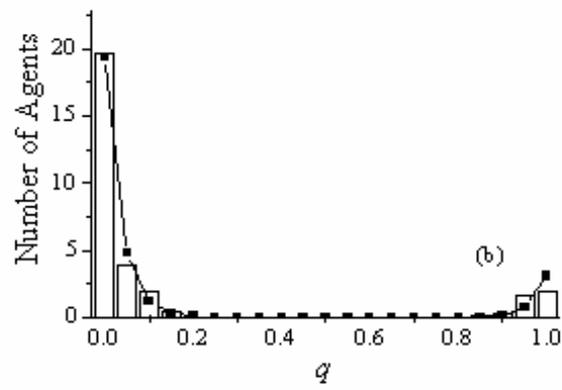

Figure 9—Di: Specialization